\documentclass[%
preprint,
 amsmath,amssymb,
 aps,
 prl,
]{revtex4-2}

\usepackage{graphicx}
\usepackage{dcolumn}
\usepackage{bm}
\usepackage{comment}
\usepackage{units}
\usepackage[T1]{fontenc}
\usepackage{xcolor}

\begin{document}

\title{Strain control of hybridization between dark and localized excitons in a 2D semiconductor}

\author{Pablo Hernández López$^{*1\dagger}$, Sebastian Heeg$^{*1\dagger}$, Christoph Schattauer$^{3}$, Sviatoslav Kovalchuk$^{2}$,  Abhijeet Kumar$^{2}$, Douglas J. Bock$^{2}$, Jan N. Kirchhof$^{2}$, Bianca H\"{o}fer$^{2}$, Kyrylo Greben$^{2}$,  Florian Libisch$^{3}$, Kirill I. Bolotin$^{2\dagger}$\\$^*$ These authors contributed to the work equally\\ $^1$ Institut f\"ur Physik und IRIS Adlershof, Humboldt Universit\"at zu Berlin, 12489 Berlin, Germany\\$^2$ Physics Department, Freie Universit\"{a}t Berlin, 14195 Berlin, Germany \\ $^3$ Vienna University of Technology\\$^{\dagger}$ Corresponding authors}

\date{
\today
}

\begin{abstract}
Mechanical strain is a powerful tuning knob for excitons, Coulomb-bound electron-hole complexes dominating optical properties of two-dimensional semiconductors. While the strain response of
bright free excitons is broadly understood, the behavior of dark free excitons (long-lived excitations
that generally do not couple to light due to spin and momentum conservation) or localized excitons
related to defects remains mostly unexplored. Here, we develop a technique capable of straining pristine suspended WSe$_2$ kept at cryogenic temperatures up to 3\% to study the strain behavior of these fragile many-body states. We find that under the application of strain, dark and localized excitons in monolayer WSe$_2$ – a prototypical 2D semiconductor – are brought into energetic resonance, forming a new hybrid state that inherits the properties of the constituent species. The characteristics of the hybridized state, including an order-of-magnitude enhanced light/matter coupling, avoided-crossing energy shifts, and strain tunability of many-body interactions, are all supported by first-principles calculations. The hybridized exciton reported here may play a critical role in the operation of single quantum emitters based on WSe$_2$. Furthermore, the techniques we developed may be used to fingerprint unidentified excitonic states.
\end{abstract}

\maketitle


\section{introduction}
Excitons are responsible for the strong absorption of light by 2D  transition metal dichalcogenides (TMDs)~\cite{Mak2010PRLa,Wang2018RoMPa,Wang2012NNa}, control the valley properties of these materials~\cite{Wang2018RoMPa,Manzeli2017NRMa,Schaibley2016NRMa}, and condense into various correlated quantum states at low temperatures~\cite{Wang2019Na,Andrei2021NRMa}. Most early studies focused on “bright” free excitons characterized by large oscillator strength and strong interaction with light fields. More recently, it became clear that other excitonic species may be critical for understanding and exploiting TMDs. First, “dark” excitons interact with light only weakly due to multiple selection rules and momentum conservation~\cite{Malic2018PRMa,Zhou2017NNa,Li2019NLa}. For example, the spin selection rules prohibit the radiative recombination of the ground state excitons in TMD systems such as WSe$_2$ or WS$_2$ composed of electron and hole wavefunctions in the K valley~\cite{Wang2018RoMPa,Malic2018PRMa,Zhang2015PRLa}. Other energetically close excitons composed of a hole wavefunction localized in the K valley and an electron in the Q or K$^{'}$ valleys are dark due to the momentum selection rule. In all cases, dark excitons feature greatly increased charge lifetime, spin lifetime, and diffusion length compared to their bright counterparts~\cite{Zhang2017NNa,Robert2017PRBa,Erkensten2021PRBa}.  These properties caused a surge of interest in such quasiparticles for storing and transporting quantum and classical information~\cite{Tang2019NCa}. The second type of excitonic species arising in TMDs other than bright free excitons are those localized by defects. Spatially localized excitonic states are delocalized in momentum space and are thought to provide the momentum needed for energy relaxation, tunneling, or emission from TMDs that is momentum-forbidden otherwise~\cite{Wu2017Na}. In addition, defect-related states function as broadly tunable quantum emitters, one of the fundamental building blocks of quantum information technologies allowing the generation of entangled photons~\cite{Toth2019ARoPCa,Wehner2018Sa,Atatuere2018NRMa, Baek2020SAa,Turunen2022NRPa,Dastidar2022Na}.

Applying mechanical strain changes the energies and the hierarchy of the excitonic states in TMDs. First-principles calculations reveal that different energy bands as well as the valleys within these bands experience varied rates of energy shift with strain~\cite{Zhang2015NLa,Aslan2018PRBa}. Excitonic species residing in these valleys inherit the corresponding energy shifts. For example, the conduction/valence band gap at the K/K$^{'}$ valleys decreases at $\sim$100 meV/\% for uniform biaxial strain \cite{Lloyd2016NLa}. Correspondingly, all excitonic species associated with these valleys including dark/bright neutral excitons, dark/bright trions, biexcitons, and their phonon replicas redshift at about the same rate. In contrast, excitons from K/Q valleys of e.g. WSe$_2$ have much weaker strain dependence, while localized excitons associated with defects remain primarily unaffected by strain~\cite{Khatibi20182Ma,Zhang2015NLa,Aslan2018PRBa}. This suggests the possibility of using the strain response signature to identify excitonic species. Even more interesting, the energy alignment between excitonic species belonging to different valleys becomes strain-dependent. As these species are brought into resonance, we expect energy transfer and hybridization between them~\cite{Linhart2019PRLa,Robert2017PRBa,Yang2019NRLa}. 

The arguments above show that mechanical strain engineering can be used to identify, generate, and tune novel hybridized excitonic states in TMDs. This possibility raises several questions. First, what are the properties of the hybridized states? They are generally expected to combine the traits of underlying excitons before hybridization. This is especially noteworthy in the case of the hybridization between bright defect-related localized excitons and free dark K/K$^{'}$ excitons featuring high oscillator strength and long diffusion length, respectively~\cite{Linhart2019PRLa,Robert2017PRBa,Yang2019NRLa}. Second, are there possible applications for the hybridized states? As strain is required for the operation of TMD-based quantum emitters, we believe that the hybridized states may be critical for the operation of these emitters and contribute to their high brightness. More broadly, the ability to manipulate the emission from long-lived dark states may prove important for various excitonic transport devices~\cite{Tang2019NCa}. Finally, how do we realize strain-induced hybridization experimentally? Observing well-separated peaks due to normally faint dark or localized excitonic species via optical spectroscopies requires clean, controllably strained devices operating at cryogenic temperatures. In contrast, most existing strain-engineering techniques either function at room temperature only~\cite{Yang2021Ia,He2013NLa,Kovalchuk20202Ma,Harats2020NPa} or do not allow in-situ control of the strain level~\cite{Lee2013NLa,Gill2015ANa,Khestanova2016NCa,Branny2017NCa,Parto2021NCa}.

Here we address these questions by implementing an electrostatic-based straining approach capable of straining a pristine suspended WSe$_2$ monolayer at cryogenic temperatures. We identify two types of excitonic species in addition to the well-known free bright excitons: i) the free dark excitons localized at K/K$^{'}$ valleys and ii) a pair of bright localized excitons related to shallow defects states. These two types of excitonic states have starkly different strain dependencies and are brought into energetic resonance at $\sim 1\%$ and $\sim 2.5\%$ strain (dependent on temperature). At resonance, we observe signatures of the formation of the new hybridized state, including orders-of-magnitude increased photoluminescence of dark excitons and avoided crossing behaviour between dark and localized excitonic species. Our data in combination with first-principles modelling suggests that the hybridized state combines the features of dark and localized excitons. Finally, some signatures of energetic resonance survive up to room temperature allowing us to observe pure defect emission. 

\begin{figure}[tp]
	\centering
	\includegraphics[width=12.5cm]{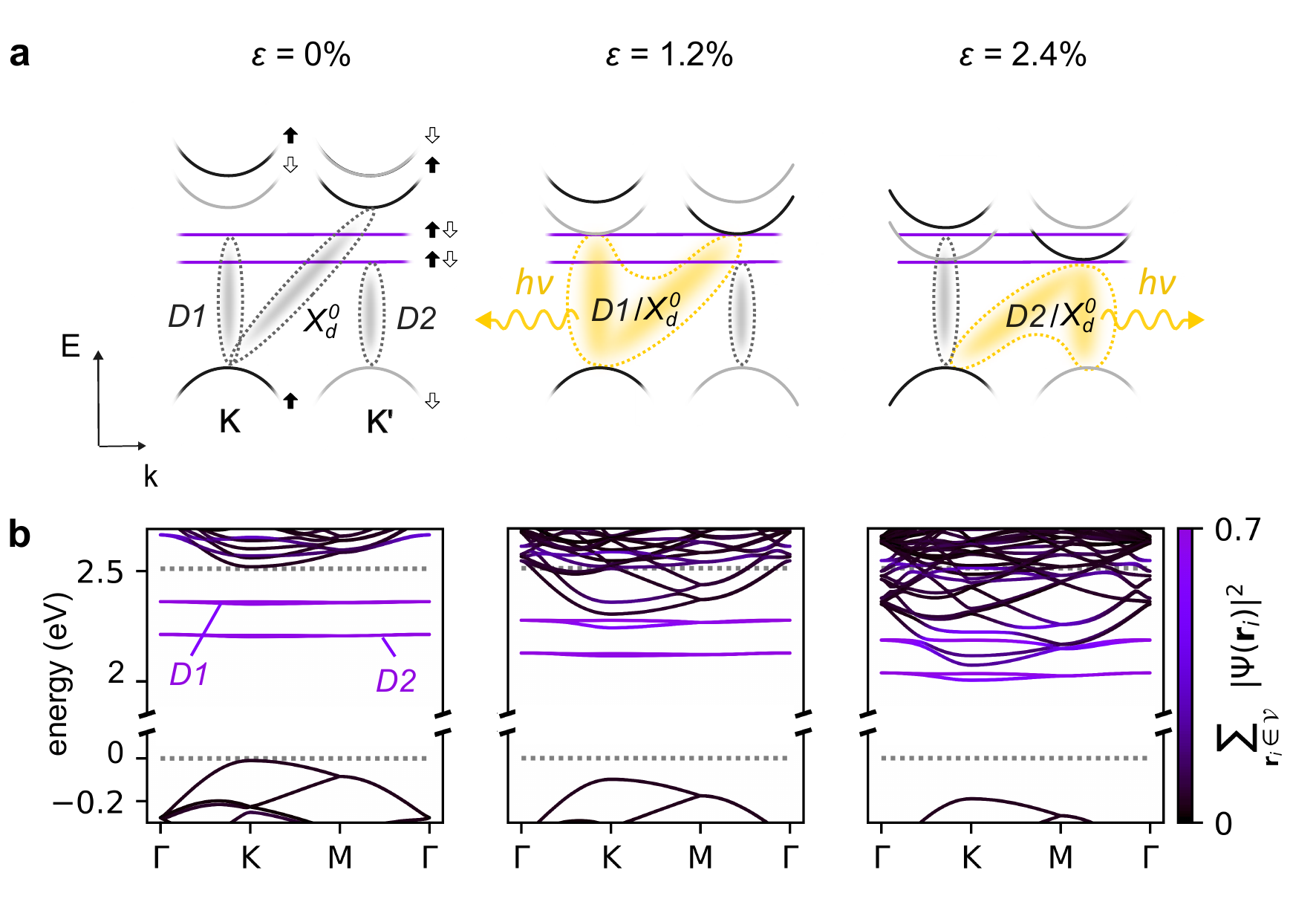}
	\caption{\textbf{Hybridization of dark exciton and defect states.} ({\bf a})~Schematic illustration of the K and K' valley bandstructure of WSe$_2$ at several different strain levels. Filled and empty arrows denote spin. ma
	({\bf b})~Corresponding bandstructure calculated via maximally localized Wannier functions generated from ab-initio calculations (see Methods for calculations details). The color scale of the bands denotes their degree of localization in the immediate vicinity $\mathcal{V}$ of the defect site. Dark excitonic states $X_d^0$ associated with the K point becomes resonant with excitons related to defect-related mid-gap states $D1$ and $D2$  at 1.2\% and 2.4\% strain. Note that the energy difference between $X^0_d$ and $D1,D2$ is determined by the built-in strain in addition to externally applied strain. Since the built-in strain is temperature-dependent due to the thermal extension of involved materials as well as other effects \cite{Plechinger2015pssRRRLa}, strain values at which the hybridization occurs also depend on temperature, as discussed in detail in Fig. 5.}
\label{Fig1}
\end{figure}

\textbf{Model of exciton strain-dependence.} In WSe$_2$, the low-energy physics is determined by the maxima of the valence bands and the minima of two spin-split conduction bands around the K/K$^{'}$ points of the Brillouin zone, Fig.~\ref{Fig1}a. The lowest-lying excitonic states are dark ($X_d^0$) , since the optical transition between the top of the valence band and the bottom of conduction band within one valley are prohibited by the spin selection rules, while the transition across valleys requires additional momentum. A point defect such as a Se vacancy in WSe$_2$ features two strongly localized (spin-degenerate) defect states ($D1$ and $D2$) within the band gap~\cite{Yang2019NRLa,Linhart2019PRLa}, slightly below the conduction band minimum. The wavefunctions of excitons associated with these states are largely formed from the valence band hole and an electron localized at the selenium vacancy (Fig.~\ref{Fig1}b). As the defect breaks translational invariance, radiative recombination of $D1,D2$ excitons is allowed. Despite the large resulting dipole matrix element, the density of isolated defect states is low. Because of that, neither defect states nor the lowest-lying excitons provide strong photoluminescence in unstrained WSe$_2$.
  
This situation changes drastically upon the introduction of strain due to contrasting strain responses of the states $X_d^0$ and $D1,D2$~\cite{Linhart2019PRLa}. Our calculations show that while the conduction band strongly downshifts with strain, the energy difference of both the valence band and the defect states remains almost constant (i.e. valence and defect bands shift at an identical, smaller rate, see Fig.~\ref{Fig1}b). Therefore, the dark exciton $X^0_d$ composed of conduction and valence band Bloch states exhibits a much stronger strain response compared to defect-related excitons $D1,D2$. As a result, $X^0_d$ becomes energy-degenerate with $D1$ at $\sim1\%$ and with $D2$ at $\sim2.5\%$ strain (Fig.~\ref{Fig1}b, top).  Following Fermi's golden rule, we evaluate the dipole transition matrix element -- including contributions from both valleys -- to estimate the resulting photoluminescence signal from an optical decay of the different possible excitonic states (see Supplementary Note 2 for technical details). We find that at resonance conditions, the oscillator strength of the dark state $X^0_d$ increases by several orders of magnitude. We interpret this striking increase in the oscillator strength as the result of hybridization between dark and defect-related excitons. The hybridized state inherits the bright character of the defect exciton, allowing an efficient de-excitation channel for dark excitons. These theoretical findings imply the possibility of continuously tuning the hybridization between excitonic species as well as their oscillator strength using strain. These phenomena are largely unstudied experimentally mainly due to the difficulties associated with reaching high values of mechanical strain at cryogenic temperatures.

\begin{figure}[tp]
	\centering
	\includegraphics[width=\textwidth]{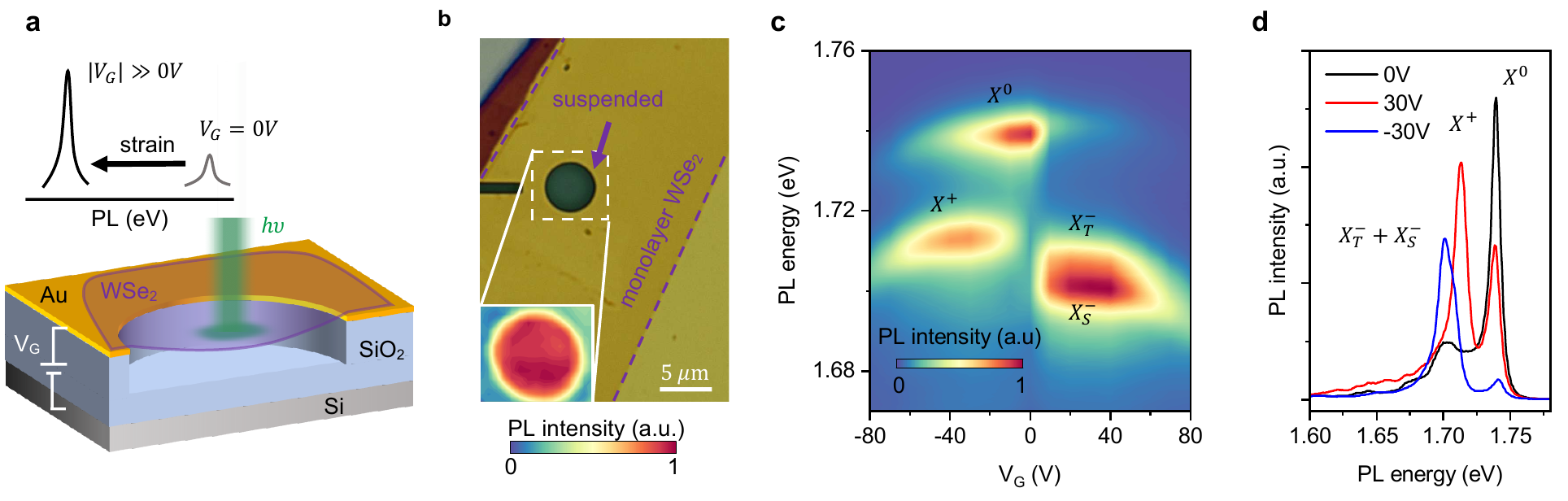}
	\caption{\textbf{Straining suspended WSe$_2$ monolayers by electrostatic forces.} ({\bf a})~Schematic of the device and measurement scheme. Applying a voltage $V_G$ between a WSe$_2$ monolayer suspended over a hole and the Si backgate mechanically strains the WSe$_2$. We probe the strain by recording PL spectra from a small region at the center of the suspended WSe$_2$ where the strain is spatially homogeneous and biaxial in nature. ({\bf b})~Microscope image of a typical device. Inset: Room temperature PL map of the suspended WSe$_2$ membrane. ({\bf c})~PL spectra as a function of $V_G$ at $T=$\unit[100]{K} taken with the laser focused at the center of the device. Selected spectra are shown in ({\bf d})~Neutral exciton ($X^0$, \unit[1.738]{eV}), positively  charged trion ($X^{+}$, \unit[1.713]{eV}), negatively charged intervalley trion ($X^{-}_{T}$, \unit[1.708]{eV}), and negatively charged intravalley trion  ($X^{-}_{S}$, \unit[1.701]{eV}) appear  at the energies close to what is reported in literature for high-quality hBN encapsulated devices ~\cite{Rivera2021NCa,He2020NCa,Liu2020PRLa,Kirchhof2022a}. Changes in intensity and peak position for $\unit[-70]{V} < V_{g} < \unit[+70]{V}$ closely resemble those encapsulated devices, showing that charging dominates the device behaviour for this regime. The near-zero $V_G$ position of the charge neutrality point and the disappearance of the $X^0$ peak outside of the $\unit[-10]{V} < V_{g} < \unit[+50]{V}$ region confirm the good optical quality of our samples.}
\label{Fig2}
\end{figure}

\textbf{Straining approach.} We solve the challenges associated with the application of controlled uniform strain to a high-quality  WSe$_2$ device at low temperature by using an electrostatic straining approach. In our technique, the strain is generated by applying a gate voltage $V_G$ between a suspended monolayer WSe$_2$ membrane and an electrode below, Fig.~\ref{Fig2}a. This approach combines several key advantages. First, electrostatic straining functions at cryogenic temperatures, unlike other straining approaches based on, e.g., bending elastic substrates or pressurizing a 2D membrane with gas that only work near room temperature~\cite{He2013NLa,Kovalchuk20202Ma,Harats2020NPa}. Second, the approach is capable of large strain values up to a few percent, higher than the limits of other techniques~\cite{Yang2021Ia,Branny2017NCa}. Such high strain is necessary to reach and surpass the predicted regime in which hybridization between different excitonic species occurs. Third, suspended 2D materials are not affected by substrate-related scattering and therefore feature high optical quality necessary to resolve closely-spaced excitonic states~\cite{Rivera2021NCa,Aslan20212Ma,He2020NCa,Liu2020PRLa}. The main complication associated with using  electrostatic forces for straining is that the carrier density inside the device changes together with the strain level. Therefore, a careful analysis disentangling the effects of doping and strain is required.  

Quantitatively, we describe the carrier density $n$ and strain level $\varepsilon$ induced in our device in response to gating via 

\begin{equation}\label{EQN:strain}
n=C_GV_G+n_0,\quad\quad \varepsilon \approx \left\{\begin{array}{cl}\alpha V_G^4 +\varepsilon_0, &  \alpha V_G^4 \ll \varepsilon_0,\\ \beta V_G^{4/3}, & \alpha V_G^4 \gg \varepsilon_0\end{array}\right..
\end{equation}

Here $C_G$ is the gate capacitance of the device, $\alpha$, $\beta$ are constants dependent on device geometry, $\varepsilon_0$ and $n_0$ are built-in strain and charge doping, respectively (mechanical details in Supporting Information Fig.~S2). These equations show that when the gate voltage is low, $|V_G| < (\varepsilon_0/\alpha)^{1/4}\sim 100$~V in a typical device, it mostly controls the carrier density while leaving the strain nearly constant. At large $|V_G|$, mechanical strain starts to change rapidly, while being symmetric with respect to the sign of $V_G$.

To realize the straining geometry described above experimentally, we suspend a  WSe$_2$ monolayer on top of a circular hole  etched into an Au/SiO$_2$/Si substrate (Fig.~\ref{Fig2}a,b). The device is loaded into an optical cryostat and measured via photoluminescence (PL) spectroscopy (See Methods for details). To establish the quality of our device and to identify the excitonic species, we start by exploring the low gate voltage regime $|V_G|<80$ V, where changes in the carrier density dominate the optical response of WSe$_2$.  For now, we focus on a temperature of \unit[100]{K} where only the most intense excitons are visible and examine lower and higher temperatures later. Fig.~\ref{Fig2}c,d shows a PL spectral map and spectra at three selected $V_G$. We observe narrow PL peaks corresponding to neutral exciton (X$^0$) and charged trions ($X^+, X^{-}_{T}$, $X^{-}_{S}$) at energies close to reported values~\cite{Rivera2021NCa,He2020NCa,Liu2020PRLa,Kirchhof2022a}. The X$^0$ peak width of \unit[7.5]{meV} is close to the $\unit[4-6]{meV}$ reported at  $T<\unit[5]{K}$ for high-quality WSe$_2$ encapsulated in hBN samples~\cite{Liu2019PRLa,Rivera2021NCa}.

\begin{figure}[tp]
	\centering
	\includegraphics[width=11cm]{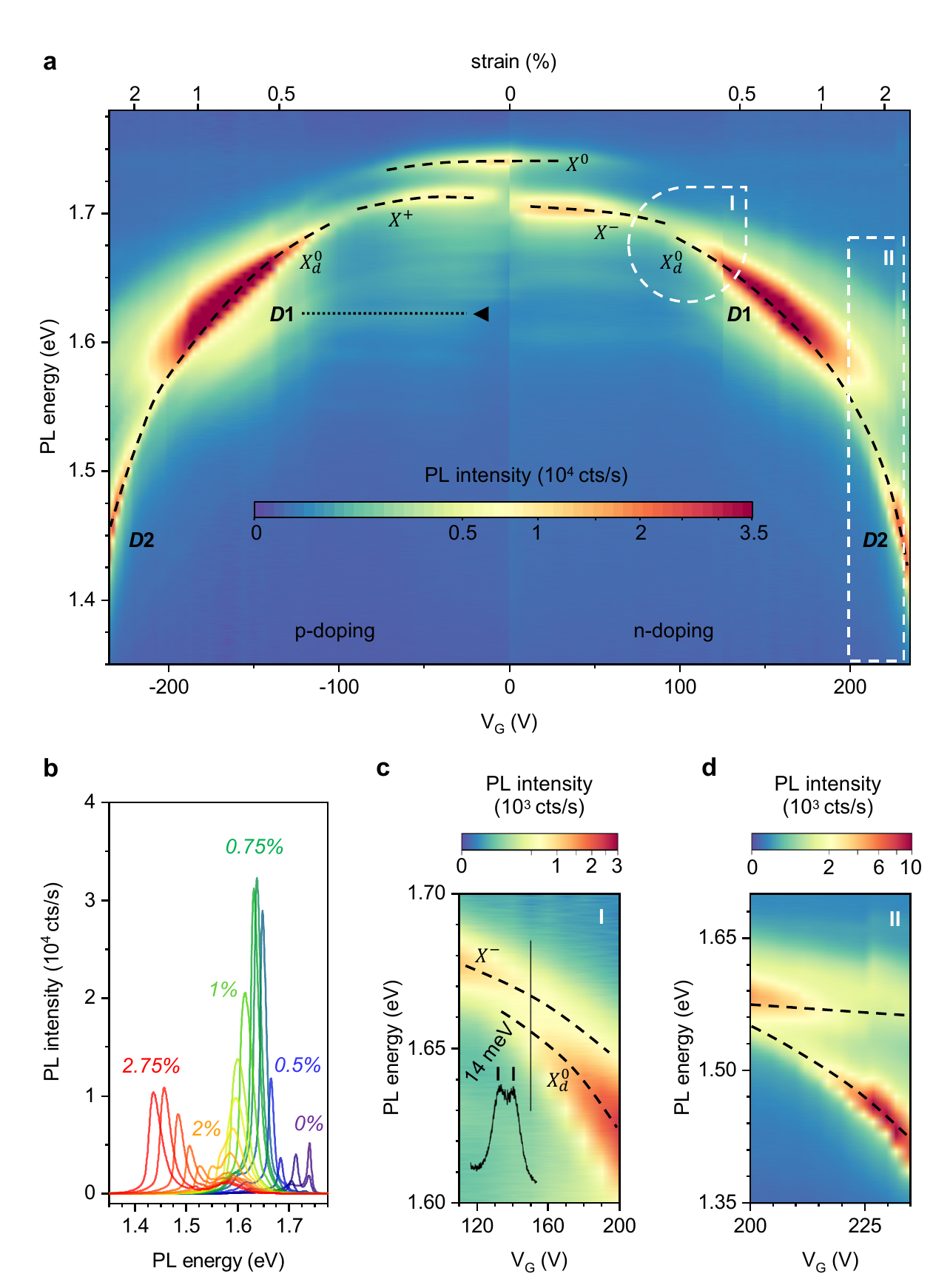}
	\caption{\textbf{Strain response of excitonic states in  WSe$_2$ at 100~K.} ({\bf a}) PL spectra map of a suspended WSe$_2$ monolayer (device 1) at $T=\unit[100]{K}$ with individual spectra shown in ({\bf b}). Free excitonic states $X^0$, $X^{+/-}, X^0_d$ red-shift with increasing strain (The dashed lines are a guide to the eye). The state $X_d^0$ develops strong intensity maxima at energies \unit[1.63]{eV} ($D1$) and $\unit[1.45]{eV}$ (D2). The strain-independent states close to the position of $D1$ that we associate with defects are marked with dashed lines.  ({\bf c}) High-resolution map of the region "I" from another device (device 2). Notice that with increasing strain, the state $X^-$ disappears, while the state $X_d^0$ arises $\sim 14$ meV below it. ({\bf d}) Detailed map of the region "II" from device 1. Notice a PL maximum near $\sim 1.4$ eV and an avoided-crossing pattern.}
\label{Fig3}
\end{figure} 

\textbf{Strain dependence of excitonic peaks.}
We now examine the behaviour of our suspended WSe$_2$ in the high voltage regime $|V_G|>80$ V, where strain effects become dominant. Figure~\ref{Fig3}a shows the evolution of PL spectra taken with gate voltages up to $V_G=\pm \unit[230]{V}$ with individual spectra shown in Fig.~\ref{Fig3}b. We observe a redshift in the energetic position of the excitons $X^0$, $X^-$, $X^+$  with increasing gate voltage. The redshift is equal  for both polarities of $V_G$. This behaviour is a well-characterized effect of mechanical strain~\cite{Conley2013NLa,Aslan20212Ma,Kovalchuk20202Ma}. In general, uniform biaxial strain in WSe$_2$ (see Methods discussing this approximation) results in a downshift of all excitonic peaks at a rate of $\unit[100]{meV}/ \%$. We use this value to fit the observed peak shifts to Eq. 2 and extract the effective 2D Young's modulus of \unit[100]{N/m}, close to previously reported values~\cite{Aslan20212Ma}. We reach strain values of almost $\unit[3]{\%}$ in our suspended WSe$_2$ devices~\cite{Aslan2018PRBa}.

\begin{figure}[tp]
	\centering
	\includegraphics[width=0.5\textwidth]{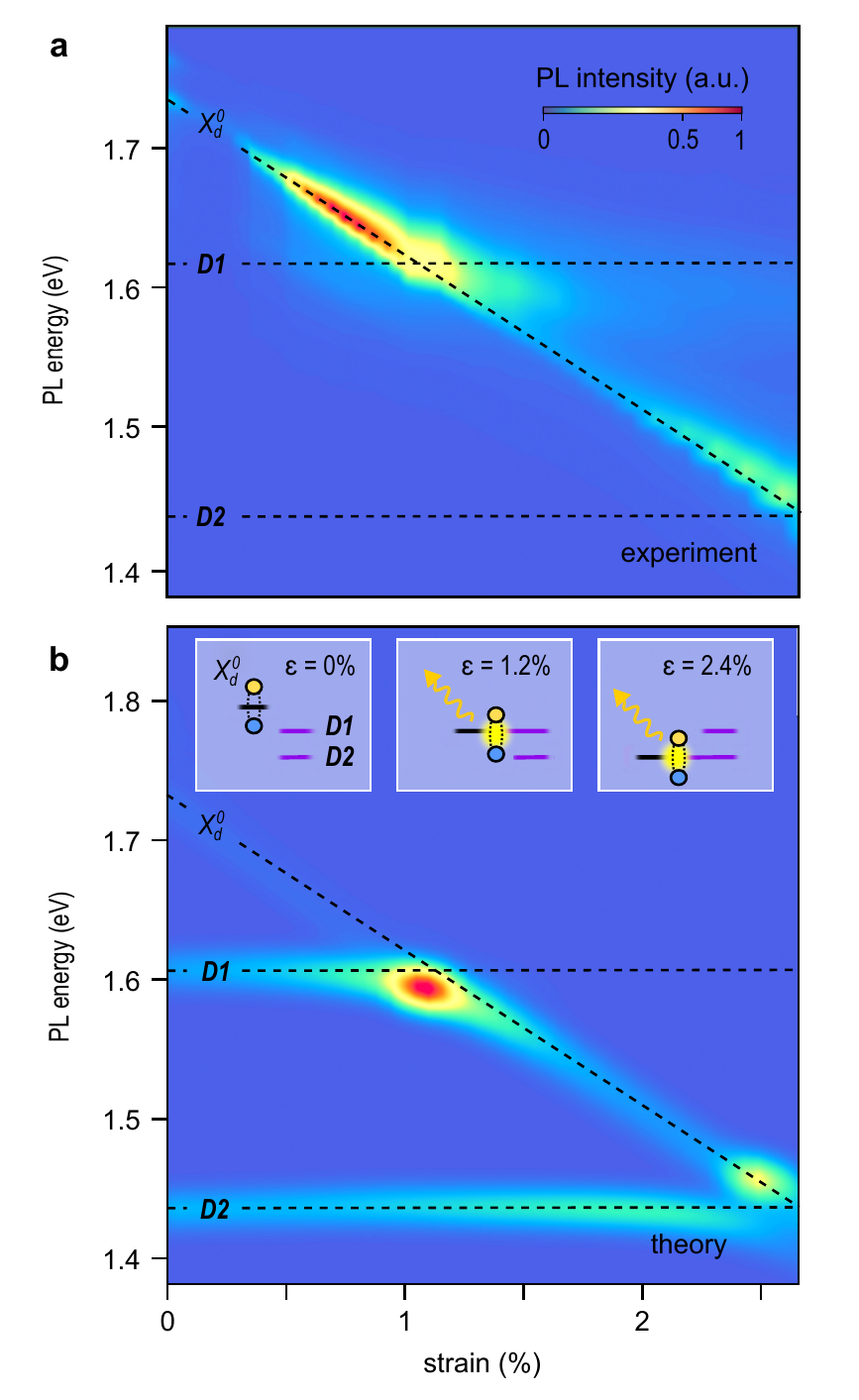}
	\caption{\textbf{Strain-related hybridized states: comparing experiment and theory.} Experimental ({\bf a}) and modelled ({\bf b}) PL spectra map of WSe$_2$ vs. strain. Different shift rates vs. strain for the defect-related states $D1,D2$ and a dark excitonic state $X^0_d$ are evident. When $X^0_d$ is energy resonant with either $D1$ (around $\varepsilon \sim 1.2 \%$) or $D2$ (around $\varepsilon \sim 2.4 \%$) a strong increase in oscillator strength and avoided-crossing type behaviours are seen. Insets show on- and off- resonant alignment between dark- and defect-related excitons. See Supplementary Note 1 for the discussion of the differences between the experiment and theory.}
\label{Fig4}
\end{figure}

We now focus on several trends in Figs.~\ref{Fig3}a and b that are not expected from a simple model based on Eq.~\ref{EQN:strain}  that only considers spectral redshift due to strain and gating-related redistribution of the oscillator strength between trions and neutral excitons. Our first and most striking observation is the highly non-monotonic evolution of the PL intensity with increasing strain. After the initial intensity drop, c.f. Fig~\ref{Fig3}a, the PL intensity increases to reach a maximum labelled $D1$ at $V_G=\pm\unit[160]{V}$ ($\sim 0.7 \%$ strain) and $\unit[1.63]{eV}$, where the integrated PL intensity is a factor $6$ higher than the combined intensities of neutral excitons and trion for $V_G=\unit[0]{V}$. It then drops and peaks again with a maximum (labelled $D2$) at $V_G=\pm \unit[230]{V}$ ($\sim \unit[2.5]{\%}$), with the peak position at $\unit[1.45]{eV}$. The state associated with greatly increased PL intensity emerges $\unit[14]{meV}$ below the trion state as the trion gradually disappears (Fig.~\ref{Fig3}c) and dominates the PL when it approaches the energies of $D1$ and $D2$. From its energy position, we identify (Methods) the state as a dark exciton, $X_d^0$~\cite{Noori20192Ma}. Note that the energetic proximity between dark exciton and trion requires high resolution measurements to resolve the two peaks.

Next, in the region $|V_G|<100~V$ we observe faint lines around 1.6 eV that are largely strain-independent (Fig.~\ref{Fig3}a, triangles), unlike other brighter excitonic peaks. In addition, these lines exhibit a saturating power dependence that contrasts with the near-linear dependence for neutral excitons and trions (see Supporting Information Fig.~S4). The saturating power behaviour is characteristic of defect-related states~\cite{Schmidt1992PRBa,Wu2017Na,Barbone2018NCa}. The near strain-independence suggests that the state does not involve Bloch states in the conduction band (Fig.~\ref{Fig1}). Finally, the comparison of the energy position of the state with our tight-binding calculations allows to unambiguously associate this state with the $D1$ exciton. Critically, the strain-independent $D1$ exciton becomes energetically degenerate with the downshifting state $X_d^0$ at about 1\% strain. The first observed PL maximum happens close to this strain value. Furthermore, the second maximum observed in PL occurs when the dark exciton is at resonance with the state $D2$ predicted from theory, at $\sim \unit[2.5]{\%}$ strain. 

\textbf{Comparison with theory.} Summarizing our discussion so far, the observed strong increase in photoluminescence occurs when a strain-related modification of the energy spectrum brings the dark exciton $X_d^0$ into an energetic resonance with defect-related states $D1,D2$. This behavior is especially striking since all of these states are either very faint or invisible in unstrained devices. The only explanation consistent with the data in Fig.~\ref{Fig3}, including changes in the intensity and the appearance of new spectral features, is strain-related hybridization between the dark exciton $X^0_d$ and defect-related states $D1$ and $D2$, as predicted by our tight-binding model. In Fig.~\ref{Fig4}, we directly compare the experimental data, now plotted as a function of strain (calculated from the shift of the excitonic peaks, see Methods), to the theoretical modelling. All of the features of the above model are evident in our data, including the linear shift of $X^0_d$ with strain, near strain-independence of $D1,D2$, and an order of magnitude increase in PL intensity when $X^0_d$ into resonance with $D1$ or $D2$. To confirm our interpretation, we have scrutinized other confounding phenomena that may contribute to our data such as strain-dependent scattering between K- and Q-valleys~\cite{Aslan2018PRBa,Aslan20212Ma} or interference effects, but find that none of these effects can explain our experimental observation (see Supplementary Note 1 and Fig.~S3).

\begin{figure}[tp]
	\centering
	\includegraphics[width=\textwidth]{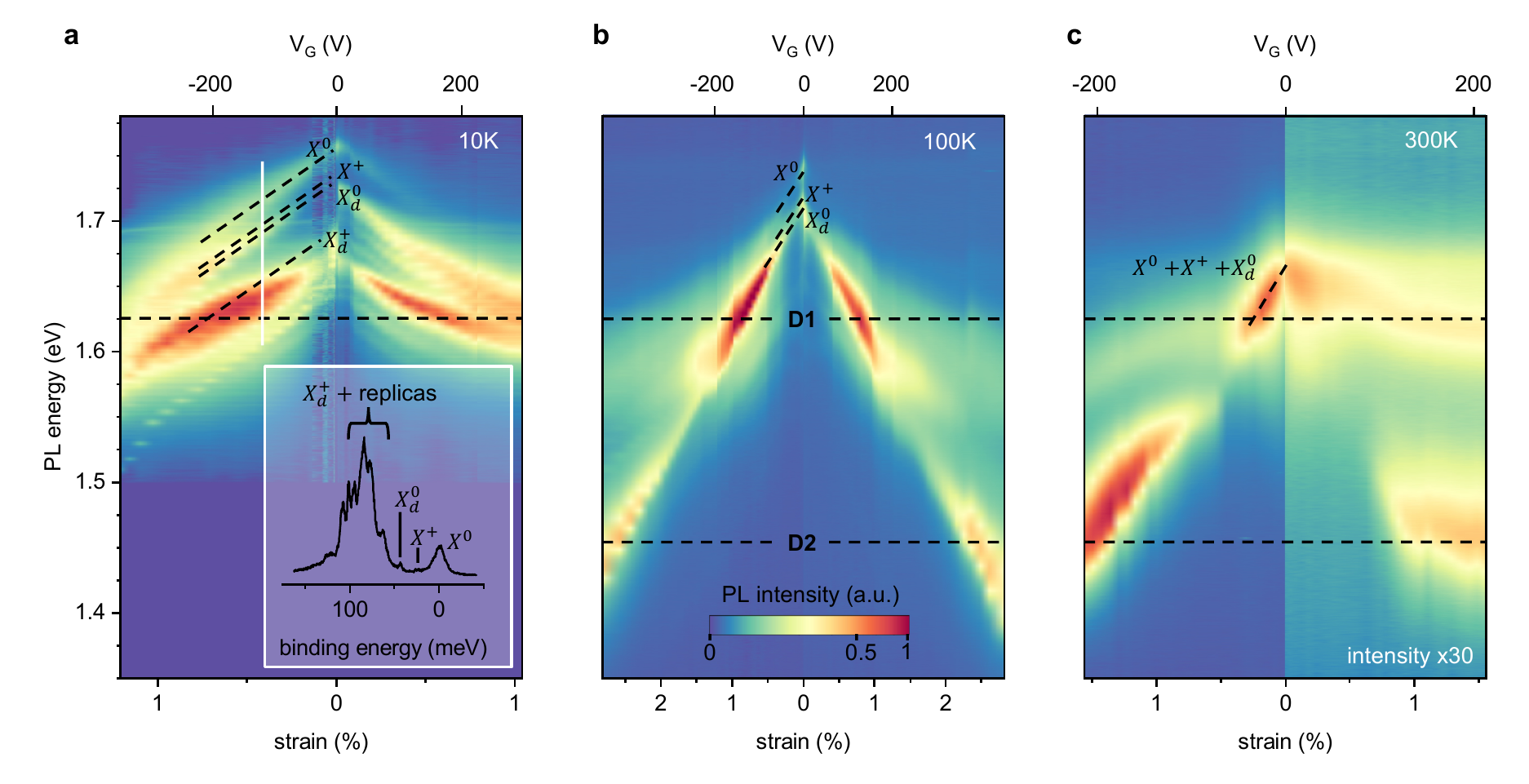}
	\caption{\textbf{Temperature-dependent strain response.} PL spectra vs. strain maps acquired at (a) T=$10$~K, (b) $100$~K, (c) $300$~K. Note that while the energy of the unstrained, free excitonic states ($X^0$, $X^+$, $X_d^0$, the corresponding spectral lines are marked with dashed lines) redshifts with lowering temperature, the features $D1$ and $D2$ associated with the corresponding defects stay roughly energy-independent. Due to the temperature-related changes in mechanical constants, only the $D1$ state is resolved at 10~K within our accessible $V_G$ range. In contrast, we resolve the hybridization with both states $D1$ and $D2$ at $100$~K and 300~K. As opposed to the $T$=10~K data in (a) that is symmetric with respect to p- and n- doping, the intensity of the n-branch at 300~K in (c) is roughly 30 times lower than that for the p-branch}
	
\label{Fig5}
\end{figure}

\textbf{Temperature-dependent data.} Having understood the strain-dependent hybridization between dark free and bright defect-related excitons, we now want to elucidate the nature of the states undergoing hybridization. For this, we turn to more complex data at temperatures between $10$~K and $300$~K  (Fig.~\ref{Fig5}). First, the strain tunability  of the excitonic states (as quantified by their energy shift within our full voltage range) decreases at low temperatures. At $V_G=200$~V, we generate 2.2\% strain at 300~K and only 0.9\% at 10~K. This difference is related to temperature-dependent changes in the built-in strain and mechanical constants of WSe$_2$~\cite{Morell2016NLa,Nicholl2015NCa}. Second, the strong enhancement in the PL intensity occurs at roughly constant energy for $T=10,100,300$~K (dashed lines in Fig.~\ref{Fig5}) despite a strong temperature-dependent red-shift in the unstrained $X^0$ and $X^{+/-}$ excitonic peaks with decreasing $T$. As discussed earlier, this behaviour is consistent with near strain-independence of defect-related excitons $D1$ and $D2$. 

Third, while at $100$~K and $300$~K the state that dominates the photoluminescence after hybridization with either $D1$ or $D2$ matches the energy position of $X_d^0$, we observe a more complex behaviour at $10$~K, Fig.~\ref{Fig5}a. A state that lies $\sim 80$~meV ($\sim 87$~meV) below $X^0$ for p- (n-) doping dominates the spectrum. This state is faint at zero strain, redshifts at the same rate as other excitonic states when the strain is increased, and its intensity reaches a maximum when it is brought into resonance with the defect-related state $D1$. The energy position of that state is within a band of states related to dark trions X$_d^+$, X$_d^-$ and their phonon replicas ~\cite{Huang2016SRa,Rivera2021NCa}. We therefore suggest that at low temperatures the strain-driven hybridization between dark trions X$_d^+$, X$_d^-$ on one hand and the defect state $D1$ on the other hand dominates the emission in our devices. The fact that such a behaviour is seen only at low temperature is likely due to the increased population of dark trions at low temperatures \cite{Arora2020PRBa}. 

Finally, at $300$~K we see a pronounced asymmetry between positive and negative voltages both in the intensity and structure of the PL spectrum, Fig.~\ref{Fig5}c, that is in contrast to the behaviour at 10~K and 100~K  (see Supporting Information Fig.~S5 for other temperatures). As the level of strain in our experimental geometry is independent of the sign of $V_G$, this asymmetry must arise from changes in the carrier density. For the n-side, the PL intensity of the n-side is about 30 times lower than on the p-side and we do not see any features associated with free excitons with linear strain dependence. We only observe weakly dispersing features at the energies of states $D1,D2$ that do not seem to undergo hybridization. We speculate that this behaviour may be related to the small oscillator strength of free excitons seen at room temperature for substrate-supported WSe$_2$ samples for n-doping~\cite{Shinokita2019AFMa,Ye2021Sa,Allain2014ANa}. We therefore suggest that the features seen on the n-side are the intrinsic contributions from $D1$ and $D2$ localized defect states that are not masked by other much stronger excitons.

\textbf{Discussion and conclusions.}
To summarize, we show that mechanical strain brings dark- and localized defect-states into resonance. When that happens, a new hybridized state with large oscillator strength is formed. That state dominates light/matter interactions in WSe$_2$. We note that the scenario we outline can in principle occur between any pair of excitonic states that have different strain dependencies, including KK and KQ excitons  \cite{Aslan2018PRBa}. Point defects other than the Se vacancies discussed here can also be involved, as long as they feature localized defect states sufficiently close to the conduction band edge to result in strain-induced hybridization. Several interesting conclusions can be drawn from our data. First, the hybridized state we observe is likely key for the operation of single quantum emitters in WSe$_2$. In these devices, highly non-uniform mechanical strain is induced in WSe$_2$ deposited onto e.g. a bed of pillars. At the point of highest strain, at the pillar's top, the hybridization conditions are fulfilled and dark excitons effectively release their energy as photons. Around the pillars' top, the strain gradient "funnels" all excitonic species, including dark, towards the point of highest strain. Since the lifetime of dark excitons is much longer than that of their bright cousins, our work shows that the area from which this funneling process "collects" the energy is much larger than what was thought previously. It is also worth noting that the processes studied here may be contributing to strain-related changes in the PL intensity previously observed in WSe$_2$ at room temperature \cite{Aslan20212Ma}. We note that similar processes may happen in other TMD materials, especially in WS$_2$, where the ground state is also dark \cite{Echeverry2016PRBa}.  Second, our results suggest the use of mechanical strain to fingerprint excitonic complexes in TMDs. The data of  Fig.~\ref{Fig3}, for example, shows that free and some defect-related excitons respond to strain very differently allowing their unambiguous identification. Finally, in the future it will be attractive to use strain engineering to "encode" charge- or spin- information into dark excitons, towards concepts such as excitonic transistors or switches.

\textbf{Methods}\\*
\textbf{Sample fabrication and measurements:} All devices are fabricated by transferring \cite{CastellanosGomez20142Ma} mechanically exfoliated WSe$_2$ monolayers onto a circular hole (diameter \unit[$\sim 5$]{$\mu$m}) etched 700nm into Au/Cr/SiO$_2$/Si (70~nm/3~nm/1500~nm/300$~\mu$m) substrates (see Supporting Information Fig.~S6). The large thickness of SiO$_2$ allows applying voltages of up to~$\pm$280~V before the dielectric rupture of the sample, but measurements are typically limited to to~$\pm$235~V for sample protection. The monolayers are many times larger than the hole to ensure device stability. Room temperature PL mapping of the finished device (Fig.~\ref{Fig2}b, inset) is used to confirm that the monolayer uniformly covers the hole area.  The devices are measured in vacuum inside a cryostat with the base temperature of 4~K. To avoid damage to the suspended WSe$_2$ during pump-down of the cryostat, a pressure relief channel is fabricated in SiO$_2$. Temperature at the sample is detected via an on-sample temperature sensor and confirmed by tracking the energetic blueshift and linewidth narrowing of the (bright) neutral exciton (Supporting Information Fig.~S5). Photoluminescence measurements are carried out in a homebuilt setup using a tightly focused laser (diameter $\sim 1\;\mu$m) placed at the centre of the device. We excite the membranes with a linearly polarized, CW laser with $\lambda=\unit[532]{nm}$ and power $\unit[\sim 10]{\mu W}$ (and in the range 100~nW -- 4.5~$\mu$W to quantify the character of excitonic states). Photoluminescence excitation spectroscopy in the range 570~nm -- 675~nm with a pulsed laser source (femtosecond Ti:Sa Chameleon Ultra II + OPO-VIS) is used to  quantify the contribution of interference effects in our data. The strain generated by applying a gate voltage between suspended WSe$_2$ and Si is largely uniform, as confirmed by modelling, mapping of the PL across the device, and high sharpness of excitonic features in strained devices. We ascribe strain values to each spectrum by measuring energetic shifts for excitonic species and assuming the shift rate $\unit[95]{meV}/ \%$~\cite{Frisenda2017n2MaAa} for all free excitons. This analysis is confirmed through separate interferometric measurements capable of directly measuring membrane displacement with nanometer resolution  (see Supporting Information Fig.~S1). To identify excitonic species in our PL data, we compare their energy position and the binding energy (measured with respect to the $X_0$ position) with detailed measurement of excitonic species in high-quality unstrained devices encapsulated in hBN~\cite{Rivera2021NCa,Li2019ANa,Liu2019PRRa,Liu2020PRLa}. While the difference in electrostatic screening between suspended and encapsulated devices leads to noticeable changes in the binding energy of the excitons, in most cases the identification is possible. 

\textbf{First-principles calculations:} We perform fully spin-polarized (non-collinear), structural and electronic optimization of the pristine cell with the DFT software package VASP \cite{Kresse1996PRBa,Kresse1996CMSa,Kresse1993PRBa,Kresse1994PRBa}. Our calculations use the Perdew-Burke-Ernzerhof (PBE) functional, include $35$~\AA~vacuum perpendicular to the membrane and a $25\times25\times1$ Monkhorst-Pack $\mathbf{k}$-space grid. Plane wave energy cutoff is set to $500$~eV and the systems are electronically converged to $\delta E\approx 10^{-9}$~eV. We find an energetically favorable lattice constant of $a_0 = 3.322$~\AA. We then generate TB parameters based on maximally localized Wannier functions (MLWFs) from a set of  DFT calculations of the pristine system and carefully interpolate TB parametrizations for intermediate strain values. Within tight-binding, we build a 17$\times$17 supercell, remove a single Se atom, and then calculate Bloch states. We evaluate optical transition amplitudes based on the dipole operator also taken from DFT (see Supplementary Note 2 for technical details).

\section{Author contributions}
S.H conceived the project. P.H.L and D.B fabricated the samples with support from B.H. and J.K. P.H.L, S.H., D.B., A.K., and S.K.,  measured the PL data assisted by K.G. P.H.L.,  J.K, D.B and A.K. performed reflectivity measurements. P.H.L., S.H. and K.B. analysed the experimental data with support from D.B. and A.K.. C.S. and F.L. performed the DFT and TB calculations. P.H.L, S.H., C.S., F.L. and K.B.  co-wrote the paper with input from all authors. K.B. and S.H supervised the project.

\section{Acknowledgments}
S.K., A.K., D.G.B., J.N.K., B.H., K.G. and K.B acknowledge support by Deutsche Forschungsgemeinschaft (DFG, German Research
Foundation) project-ID 449506295 and 328545488, TRR227, SfB951, and ERC Starting grant no. 639739. P.H.L., and S.H. acknowledge funding from the Deutsche Forschungsgemeinschaft (DFG) under the Emmy
Noether Initiative (HE 8642/1-1). C. S. acknowledges support as a recipient of a DOC fellowship of the Austrian Academy of Sciences. Numerical calculations were performed on the Vienna Scientific Cluster VSC4. 

\bibliographystyle{naturemag}
\bibliography{Strain_by_backgate_20220711}

\end{document}